\documentclass[11pt]{article}
\usepackage[utf8]{inputenc}

\usepackage{url, graphicx, natbib, xcolor, amssymb}
\usepackage{multirow}
\usepackage{array,longtable} 
\usepackage{boldline} 

\usepackage[tableposition=t]{caption} 
\usepackage{setspace}
\DeclareCaptionFont{singlespacing}{\setstretch{1}}
\usepackage[margin=1in]{geometry}

\usepackage{footnote}
\usepackage[hang]{footmisc}
\usepackage{hyperref}
\hypersetup{
    colorlinks=true,
    linkcolor=blue,
    citecolor=blue,
    urlcolor=cyan,
}

    \makeatletter
\def\@fnsymbol#1{\ensuremath{\ifcase#1\or \bigstar\or 
\dagger\or \ddagger\or
   \mathsection\or \mathparagraph\or \|\or **\or \dagger\dagger
   \or \ddagger\ddagger \else\@ctrerr\fi}}
    \makeatother

\title{\textbf{Making Science Personal: Inclusivity-Driven Design for General-Education Courses}}
\author{Christine O'Donnell$^{1}$\thanks{\hspace{0.5em}Email: \url{caodonnell@email.arizona.edu}}\hspace{0.75em}, 
Edward Prather$^{1}$, \& Peter Behroozi$^{1}$
\\
\footnotesize{$^{1}$ \textit{Department of Astronomy and Steward Observatory, University of Arizona, Tucson, AZ 85721, USA}}}
\date{}

\begin{document}

\maketitle


\begin{abstract}
    General-education college astronomy courses offer instructors both a unique audience and a unique challenge. For many students, such a course may be their first time encountering a standalone astronomy class, and it is also likely one of the last science courses they will take. Thus, in a single semester, primary course goals often include both imparting knowledge about the Universe and giving students some familiarity with the processes of science. In traditional course environments, students often compartmentalize information into separate ``life files'' and ``course files'' rather than integrating information into a coherent framework. The astronomy course created through this project, taught at the University of Arizona in Spring 2019, was designed around inclusivity-driven guiding principles that help students engage with course content in ways that are meaningful, relevant, and accessible. Our course bridges the gap between students' ``life'' and ``course files'', encourages and respects diverse points of view, and empowers students to connect course content with their personal lives and identities. In this paper, we provide insight into the guiding principles that informed our course design and share research results on the effectiveness of the instructional strategies and assessment techniques implemented in the course.
\end{abstract}

\section{Introduction: General-Education College Curricula}
\label{sec:edpaper_intro}
Many universities require students to take general-education courses spanning science, history, writing, etc. At the University of Arizona, the curriculum's goals\footnote{\url{https://catalog.arizona.edu/policy/general-education-curriculum}. We note that the University of Arizona's goals are not dissimilar to other institutions' general-education goals.} include ensuring that all students have foundational knowledge from subjects beyond their major so that they can appreciate how their discipline fits into and supports a broader societal context. Additionally, the curriculum aims to encourage acceptance of people with different backgrounds and give students a ``deepened sense of self''. In the sciences, general-education courses often aim to impart both discipline-specific knowledge and science practices/skills such as critical thinking. 

However, as \cite{Fink} argues, students often compartmentalize course content into a ``course file'' for homework/tests or a ``life file'' for use in their everyday lives. We believe general-education courses need to bridge the gap between these files. Thus, a ``significant learning experience'' that empowers students to connect or add something to their ``life file'' will create lasting learning. These experiences can feature 
\begin{enumerate}
    \item Integrating course content with other disciplines or aspects of life, which directly addresses the general-education goal to enable students to grapple with society's complex interdisciplinary issues; building these connections can guide students to understand the relevance that science already has in their lives.
    \item Focusing on the human dimension to encourage students to learn more about themselves and others, which addresses the human story and affective domain of learning \citep{Krathwohl64} so that students can gain a greater appreciation of people from diverse backgrounds and build stronger self-identities. 
\end{enumerate}

One approach for science courses to address identity is a ``worldviews'' approach \citep{Cobern96}. While science is an integral component of technology, policy, and everyday life, the ``public alienation'' from science instead makes it into a disconnected subject. Cobern argues that science needs to be taught jointly with other disciplines to create a ``coherence view of knowledge'' so that students will view scientific concepts as ``superior'' (either in terms of usefulness or power) to their pre-instruction conceptions. However, the worldviews approach has a limited ability to create significant learning experiences. For example, by implying that certain concepts are ``superior'', it can reinforce systems against students from marginalized backgrounds rather than valuing students' diverse experiences. Additionally, the worldviews approach focuses solely on sociocultural identities and ignores personal identities.

A course that addresses both sociocultural and personal contexts will access more learning dimensions and can create a more welcoming environment. Science has traditionally been taught as being a ``neutral'' or ``acultural'' topic. However, science represents a culture unto itself that has been shaped by and for dominant groups, and this culture can drive away those from non-dominant backgrounds \citep[e.g.,][]{NRC_00, NRC_09, Seymour97, Brickhouse01, Brown05}.  By addressing the interplay between students' existing (and developing) identities, larger sociocultural framings, and science's culture, we can create a more inclusive environment that is welcoming of diversity \citep{Reveles08, Carlone07}. Rather than reinforcing the idea that students have to assimilate into science's culture, we can encourage participation by guiding students to see science as part of and valuable to their own identities \citep{NRC_07}.

In this paper, we present new inclusivity-driven classroom instructional strategies that attend to students' identities, and our research assesses whether this curriculum leads students to integrate their ``course files'' with their ``life files''. Some education research has explored equity in the college classroom, e.g., related to gender \citep{Weinburgh95, Roychoudhury95} or students with disabilities \citep{Norman98, Bell02}. However, many of these studies focus on student grades (``achievement gaps''), whereas we focus on assessing students' experiences and connections to their identities. Below, we first discuss the unique nature of general-education astronomy courses, followed by a description of our specific course. We present guiding principles of our course design, examples of course content, and assessment results. Our work offers a framework from which instructors can build an inclusive mindset into their own courses that ``engage[s] students in learning that is meaningful, relevant, and accessible to all'' \citep{Hockings10}.

\subsection{Astronomy General-Education Courses}
\label{sec:edpaper_astro_gened}

Non-science majors often take astronomy to fulfill general-education science requirements. Annually, over 250,000 students enroll in an astronomy general-education course in the US, and they represent all demographic backgrounds \citep{Rudolph10}. For many of these students, it may be both the first time they will encounter astronomy as a standalone course and simultaneously the last time they will formally engage with any science. This presents a unique challenge for instructors: they have to (1) introduce students to astronomy content and (2) address that this may be the last time our future voters, educators, etc. experience science. Previous research has investigated the teaching and learning of astronomy content through active learning strategies \citep[e.g.][]{Prather09_activelearning, Prather09_teaching} and implementing a worldviews approach \citep{Wallace13}, but they do not address students' personal identities and lived experiences.

\section{Course Background}
\label{sec:edpaper_coursebkgd}
In Spring 2019 at the University of Arizona, a team of 
\begin{enumerate}
    \item a general-education astronomy course instructor (an assistant professor in the Astronomy Department),
    \item a graduate teaching assistant (an Astronomy \& Astrophysics Ph.D. candidate), and
    \item an astronomy education researcher (a professor in the Astronomy Department),
\end{enumerate}
reformed a general-education introductory astronomy course (ASTR 201: Cosmology). The course has no prerequisites, and this was the instructor's first time teaching it, though the course itself has been offered for over a decade. 

Forty-one students enrolled. Only six students (14.6\%) intended a STEM-related major (e.g., biology or engineering), and the most commonly intended majors were business-related (11 students; 26.8\%). Nine students (22.0\%) were first-year students, twenty-one (51.2\%) were second-years, six (14.6\%) were third-years, and five (12.2\%) were fourth-years\footnote{Compared with \cite{Rudolph10}, our year distribution has fewer first-year students than is often seen in a general-education astronomy course, but our course was a ``Tier II'' general-education course which attracts a greater percentage of non-first year students. For more information on the Tier II designation, see \url{https://catalog.arizona.edu/policy/general-education-tier-one-and-tier-two}.}. Thirty-four students gave us informed consent to collect their course data for our research. 20 students responded to a short-answer self-identification prompt. Half of these students identified as female, and half identified as male; fifteen (75\%) identified as White and/or Caucasian, two (10\%) as Latino\footnote{This was the identification terminology provided by the students.}, and two (10\%) as Native American.

\begin{figure}
\centering
\includegraphics[angle=0,width=\columnwidth]{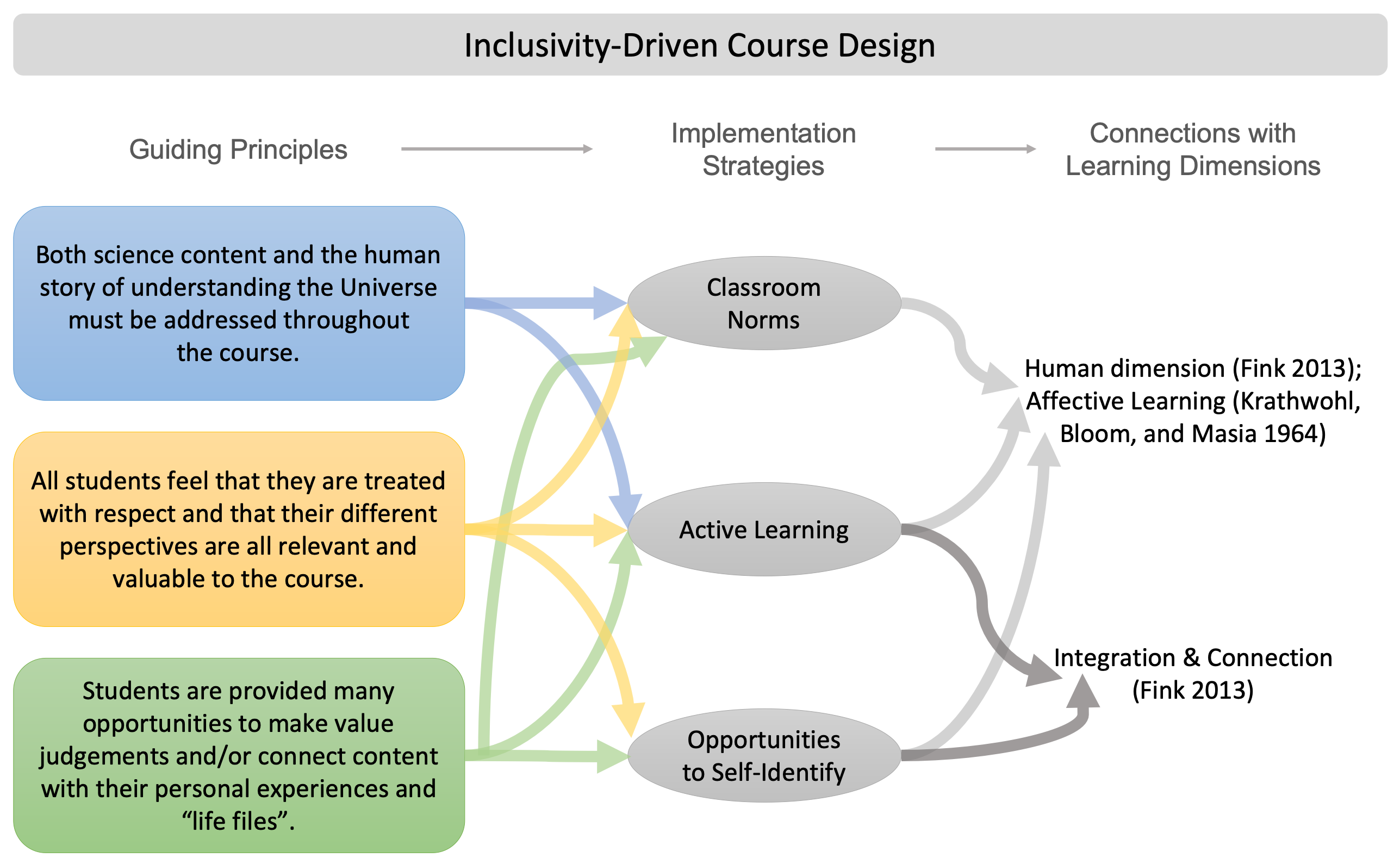}
\caption{Schematic diagram of our inclusivity-driven course design, including our guiding principles, examples of research-based implementation strategies, and connections with learning dimensions. The arrows are intended to be suggestions for implementation and are not exclusive, e.g., our guiding principle for covering both science content and the human story should not be thought of as completely absent from opportunities to self-identify.}
\label{fig:education_schematic}
\end{figure}

Our novel inclusivity-driven course design aims ``to engage students in learning that is meaningful, relevant, and accessible to all'' (Hockings 2010). We built our course around these guiding principles (Fig.~\ref{fig:education_schematic}):
\begin{itemize}
    \item Both science content and the human story of understanding the Universe must be addressed throughout the course.
    \item All students feel that they are treated with respect and that their different perspectives are all relevant and valuable to the course. 
    \item Students are provided many opportunities to make value judgements and/or connect content with their personal experiences and ``life files''.
\end{itemize}
Our course represents a pilot test of these principles.

\section{Classroom Norms}
\label{sec:edpaper_norms}

An important aspect of our design was our classroom norms. In many classes, ``norms'' are limited to established policies around grading, late assignments, attendance, etc. However, as \cite{Tanner13} describes, ``norms'' can refer more broadly to behaviors and attitudes, such as ``Everyone here has something to learn.'' To successfully establish a norm, an instructor has to not only state it but also enforce it throughout the semester.

We established a norm to acknowledge and value diverse perspectives in a way that affirmed the importance of students' lived experiences. Courses typically do not include readings or discussions on topics relevant to members of underrepresented groups \citep[e.g.,][and references within]{Harper09}. Without making intentional choices to incorporate diverse voices into the classroom, curricula that focus on dominant Western perspectives represent a form of power that implies that beliefs from different cultures are not valued \citep{Delpit88, Banks10}, which is contrary to our guiding principles.

To achieve this norm, our course explicitly acknowledged additional voices. On the very first day of the course, after a ten-minute course content overview, a member of the local Tohono O'odham Native American Nation gave a 1-hour lecture on their cultural beliefs of the Solar System, Milky Way, and other celestial objects. He also described the importance of certain days of the year, such as the solstices. This lecture tied into the course's first unit about human and cultural connections to the sky (e.g., for navigation, agricultural practices, etc.) for many different cultures (e.g., European, Egyptian, and Asian). The norm was reinforced throughout the semester through stories describing the human endeavor of science. We shared life stories of scientists, such as Cecilia Payne-Gaposchkin, an astronomer who first proposed that the Sun is composed of hydrogen and helium, contradicting the dominant theory of the time, and she faced many systemic and institutional barriers. 

\section{Active Learning}
\label{sec:edpaper_activelearning}

Active learning is not new to introductory astronomy general-education research. \cite{Prather09_activelearning} showed that active learning can significantly increase students' astronomy content learning gains. Think-pair-share activities were used to cultivate students' critical thinking \cite[e.g.,][]{Tanner13, Supiano18}. Furthermore, we adapted think-pair-share questions to incorporate inclusivity and empower students to connect with their ``life files''. 

For example, after grading homework assignments, the graduate teaching assistant reported common student struggles to the instructor, and the instructor debriefed those struggles in class. After a particularly difficult assignment, which dealt with complex math and equations as well as visualization of light bending around a black hole, the instructor led a debrief to help students connect with the enterprise of science. These think-pair-share prompts framed the debrief:
\begin{enumerate}
    \item The instructor asked students to consider all the skills they feel are helpful to do science.
    \item He then had them pair up and share/compare their sets of skills.
    \item He had the student in each pair whose name came first alphabetically share the pair's discussion. This sharing method was chosen to promote inclusion: by assigning a ``reporter/sharer'' based on a random characteristic, we provided opportunities for verbal participation by students who may not otherwise volunteer. Additionally, by choosing a random personal characteristic, we encouraged a collaborative community among our students \citep{Tanner13}.
    \item The instructor typed responses into a lecture slide, making students' ideas visible to the whole class and acknowledging each response. Student responses included open-mindedness, communication, critical thinking, creativity, and leadership.
\end{enumerate}

The instructor explicitly noted that these responses are all ``skills'', meaning that one can change them over time. Additionally, he noted that science is often done in collaborations, such as the large team that detected gravitational waves, a topic from the prior week. He stated that science is inclusive of people who can lead well but are not especially curious, people who are creative but are weaker with leadership skills, and people who can communicate and connect people. No single person has all of the skills that the students reported, and he stated that ``there's places in science for all different kinds of people with all of these different kinds of skills.'' In the authors' experiences, other science courses may emphasize a specific set of skills as being ``keys to success''. Instead, in our course, the instructor had the students create a list of skills and left it up to each student to reflect on how their own existing skills fit within science and beyond.

\section{Opportunities to Self-Identify}
\label{sec:edpaper_selfidentify}

We also provided regular opportunities for students to express their personal opinions as part of assignments and quizzes. Many studies emphasize the importance of connecting content with students' lives \citep[e.g.,][]{NRC_00, NRC_09}, and they also demonstrate the positive effects of these experiences. For example, \cite{Hulleman09} studied writing prompts in a ninth-grade science course that asked students to summarize course content and encouraged students to make connections with their lives. They found increases in both interest and course grades among students with low success expectations.

In our course, almost all class sessions included a 5-minute writing prompt that reflected on concepts from that day's lecture; students wrote responses on index cards, and a thoughtful response received full credit. For example, one topic was dark matter, which does not interact with light and therefore cannot be directly observed, but its presence can be inferred from gravitational interactions with visible matter. The corresponding writing prompt intentionally asked what students believe in but cannot see. Some responses were scientific, such as gravity or oxygen, but over 40\% of responses connected to ``life files'', such as God, souls, or love. 

Some writing prompts were expanded into homework and/or quiz questions. A unique course theme was ``reference frames'': depending on how you define your perspective, the same physical system can appear very different. For example, in our reference frame on Earth, planet orbits show unusual behavior including temporary reversals of their apparent direction (i.e., retrograde motion). However, in a reference frame in which the Sun is at rest, the planets always travel in the same direction. The choice of reference frame (i.e., the choice of coordinate system) by definition does not affect accuracy for predicting planetary motion due to gravity, which we demonstrated with orbital simulations. Nonetheless, as above, it has a profound effect on apparent motion. We made an analogy between these reference frames and having a disagreement with another person due to differing perspectives. Students had a 5-minute in-class prompt to describe a time in their lives when two opposing views were valid. The next homework asked students to write about a memorable disagreement they had with another person, what arguments supported their own view, what arguments supported the other person's view, and how the other person could rationally come to that viewpoint. Student responses on both assignments included politics (e.g., gun control, death penalty, immigration), religion (e.g., the existence of God), conflicts with family and friends, and personal topics (e.g., musical preferences). These assignments empowered students to create connections between course content and their personal lives. During debriefs, the instructor affirmed that feelings of discomfort when dealing with such questions are natural. 

\section{Student Reflections \& Assessment}
\label{sec:edpaper_results}

We report student reflections, course scores, and survey results.

\subsection{Student Reflections}
\label{sec:edpaper_reflections}

The final exam included a question that asked students about whether this class changed the way they think about their own lives or their place in the Universe. Additionally, some students provided comments in the University of Arizona's Teacher-Course Evaluations (TCE). Table \ref{tab:edpaper_reflections} reports relevant responses; all student comments on course design elements described in this paper were positive.

\subsection{Course Scores}
\label{sec:edpaper_grades}

While our course design was motivated by a desire to be inclusive of our students' personal identities, we also are sensitive to the fact that grades are an important aspect of students' motivations and course experiences. Many STEM education studies have identified grade differences across demographic identities. Table \ref{tab:edpaper_grades} summarizes the average cumulative course scores for the students that responded to our demographics survey (Sec. \ref{sec:edpaper_coursebkgd}). We observe no differences in average scores across gender ($p = 0.877$ from a Welch's $t$-test) and culture/ethnicity ($p = 0.915$). 

\subsection{Pre- and Post-Course Survey}
\label{sec:edpaper_TSSI}
Finally, we conducted a pre- and post-course survey on students' views of science. We selected 25 items\footnote{The full TSSI sample includes 60 prompts; our subset was chosen based on alignment with our courses' goals.} from the Thinking About Science Survey instrument \citep[TSSI;][]{Cobern_TSSI}, which is aligned with our goal of connecting science to students' lives. However, we adjusted the coding for four survey items. Cobern scored the survey to assess how strongly students agree with the public perception of science portrayed by scientists, educators, and journalists that associates science with properties such as superiority and exclusivity. For example, Cobern lists the item ``A person can be both religious and scientific'' as having reverse polarity, i.e., a student that believes in this public portrayal of science will respond with ``strongly disagree''. Our course affirms that there are many different yet equally valued sources of understanding, so we do not reverse the scoring of this item. An additional limitation is that the TSSI focuses on students' views of science's sociocultural context, but our course design also acknowledges personal contexts.

Table \ref{tab:edpaper_TSSI} details our survey items and results. 13 students responded to both the pre- and post-course survey. For each of the items, over 50\% of participants gave the same response across the two surveys, meaning for almost all items, differences are due to only 1 or 2 students' responses. We also note that the pre-survey averages tend to be favorable to our goals; this has been observed in other studies and makes it difficult to clearly attribute changes to an instructor's efforts and/or course design \citep[e.g.,][]{Adams13, Perkins05, Wallace13}. Comparing students' average pre-course full survey score and average post-course survey score, we see a small positive change ($\Delta = 0.07$) with $p = 0.170$ from a Wilcoxon Signed-Rank Test.

\section{Discussion \& Conclusion}
\label{sec:edpaper_disc_conclu}

Fig.~\ref{fig:education_schematic} summarizes our inclusivity-driven course design. Our course was built on guiding principles that (1) emphasized both science content and the human story of understanding the Universe, (2) respected diverse perspectives, and (3) provided students with many opportunities to make connections between course content and their ``life files''. We wove these principles through all aspects of the course, including explicit classroom norms, lecture content, in-class writing prompts, and homework assignments and quizzes. We created unique opportunities for students to share their personal thoughts, beliefs, and experiences to directly connect their own lives with astronomy content and science practices. Furthermore, we enhanced evidence-based active learning methods to improve inclusivity. For example, we introduced think-pair-share prompts that asked students to critically reflect on skills that are useful in science. In class, we explicitly emphasized that science is done by different people who each contribute different and unique perspectives and skill sets. Finally, we enhanced our think-pair-share exercises by using sharing methods that encouraged participation from all students, such as assigning a random member of each pair to report their discussions.  

The feedback from all aspects of the class, including powerful student reflections, equal course scores from different demographic groups, and overall positive responses to survey items, demonstrate that our design provided a positive experience to help students meaningfully connect science to their personal identities and ``life files''. We believe that our results are from the manifestation of our guiding principles throughout all aspects of the course, creating truly significant learning experiences that are ``meaningful, relevant, and accessible to all'' students \citep{Hockings10}. We intentionally included material that represents diverse voices, such as having a member of the local Tohono O'odham Native American Nation give the first lecture, and we discussed the nature and practices of science. These course aspects acknowledged the culture of science as well as provided students with opportunities to reflect on how their own personal lived experiences can be welcome and valued in science.

Our guiding principles provide a framework for future course iterations as well as for other instructors who wish to incorporate an inclusivity-driven mindset into their courses. Our course instructor found that implementing these principles required minimal extra work beyond what is normally required for designing a class. In fact, he found that thinking about which aspects of a topic are most relevant to students' personal lives helped a great deal to decide which material was crucial for students to take away from the course versus which material was less important. Furthermore, it was an eye-opening experience for the instructor to learn about how other cultures view astronomy. For example, the speaker from the local Tohono O'odham Native American Nation seamlessly glided between stories of creation and stories about peoples' lives on Earth, in part because the Earth and sky are literally sewn together in their view. Their culture has less ``distance'' between astronomy content and people's lives, i.e., the two realms have a high degree of overlap and relatedness. By intentionally bridging ``course files'' and ``life files'', we developed a class that was both sensitive to the dimensions of significant learning as well as our students' different cultural perspectives.

Finally, we consider several possible directions for future research.
\begin{itemize}
    \item We could improve the assessment and evaluation of the course, such as (1) a structured qualitative analysis to assess student responses throughout the semester, (2) an improved quantitative analysis by deploying a survey instrument more closely aligned with our goals, and/or (3) an in-class observational analysis to assess classroom equity (e.g., which voices are represented).
    \item We can incorporate additional course elements, such as group projects to encourage community building. These elements would access more dimensions of significant learning and provide more opportunities (1) for students to learn from one another and (2) for creating connections between content and personal experiences. 
    \item Finally, we could reform an undergraduate majors course using our design model to investigate inclusivity in these STEM-specific learning environments; this research could also examine the retention of underrepresented populations.
\end{itemize}
Given our students' feedback, our model empowers students by letting them make science a part of their identities, values their ideas and experiences, and creates a more inclusive classroom environment that can reach a broader student audience.

\section*{Acknowledgements}
We would like to thank the students who enrolled and participated in our class. Their engagement and excitement are what motivates us to pursue these research questions. We would also like to thank Dr. Lisa Elfring, Dr. Joanna Masel, and fellow participants in our Faculty Learning Community for input on our course design; Dr. Julie Libarkin and her research group for helpful comments on the paper; and Prof. Camillus Lopez from the Tohono O'odham Native American Nation. CO's graduate teaching assistantship was supported from teaching reform project funding by the Howard Hughes Medical Institute and the Accelerate for Success program at the University of Arizona. Finally, we thank the anonymous reviewers for their helpful comments to improve our manuscript.

\newpage

\clearpage

\begin{center}
\footnotesize
\setstretch{1.15}
\begin{longtable}{p{0.19\columnwidth}| p{0.3\columnwidth} | p{0.3\columnwidth}}
\caption{Verbatim student responses from a final exam question that asked students how the course changed the way they think about their own lives or their place in the Universe, as well as comments from the anonymous University of Arizona Teacher-Course Evaluations (TCE). \newline [Note: This table is split over 3 pages.]}\label{tab:edpaper_reflections} \\

\hlineB{2} \multicolumn{1}{c|}{\textbf{Topic}} & \multicolumn{1}{c|}{\textbf{Final Exam Responses}} & \multicolumn{1}{c}{\textbf{TCE Responses}} \\  \hlineB{2}
\endfirsthead

\multicolumn{3}{c}%
{{\bfseries \tablename\ \thetable{} -- continued from previous page}} \\
\hlineB{2} \multicolumn{1}{c|}{\textbf{Topic}} & \multicolumn{1}{c|}{\textbf{Final Exam Responses}} & \multicolumn{1}{c}{\textbf{TCE Responses}} \\  \hlineB{2}
\endhead

\hlineB{2} \multicolumn{3}{r}{{\textit{Continued on next page}}} \\ \hlineB{2}
\endfoot

\hlineB{2}
\endlastfoot

Classroom Norms & & It was engaging and interesting and the professor cares about everyone's thoughts and opinions on subjects. \newline\newline
The questions you guys asked allowed for honest responses, and the way they were worded made me feel comfortable expressing my actual opinion on the topics discussed! The teaching style for this class was definitely in my top three, and this is my second degree and sixth year in college so there's a biiiiig pool.\\\hline

Active Learning & & Really included people in discussions and invited questions. Very respectful professor who truly cares about his students' learning.\newline\newline
[...] his methods of questioning and getting us to think about our answers and why we chose them helped me understand not just the facts but how we got them \\ 

Opportunities to Self-Identify (e.g., writing prompts) & 
This course change my thinking. I learned how to use critical and scientific thinking to solve the problem. [...] So when we have argument, I will try to think as other people which will help me consider two or more critical thinking.\newline\newline
[...] this class has allowed me to think more critically and have an open mind. Doing the homework, and comparing astronomical concepts to things on earth helped me to think about things in a different way. I feel that when approaching problems now, I can think of many different ways to solve it.\newline\newline When we were learning about parallax and perspective, I was dealing with some family problems that have a lot to do with viewpoints. I had sat around that week on the phone, trying and trying to handle everything and get my family to understand why they are so incredibly mistaken about an issue they remain misinformed about, to the detriment of a cousin going through a rough time. We had been arguing unproductively for almost a month, and then we learned about how perspective changes how we receive information. Taking that and applying it to the conversation, my cousin and I managed to make them understand why she chose what she did and while unhappy, they accepted it. I apply this to most discussions now, and I've become a better advocate because of it.&
I did like the writing activities we had for each class where a question was posed that we would write the answer to such as ``Think of a time when... happened to you'' or the like \\ 

Other Comments Related to Students' Attitudes & I used to think about the universe in a fearful way, and I think I've managed to get over that quite well, \underline{because} I know more about it now. \newline\newline
I feel more solid about my view of the universe as the ``divine'' (for lack of a better word) after this class. The reason I see it that way is because divinity is supposed to be beautiful, omnipresent, omniscient, mysterious. Earth is like a mini universe, and so is the Solar System, the Galaxy, the Quadrant, etc. Even our cells are tiny collections of cosmic dust. Just because I don't believe in a conscious deity doesn't mean I don't find the concept in the universe. Learning about the different celestial bodies and forces and how gravity is not really a force (which, that blanket analogy is told to everyone now), seeing it all come together is as close to divinity as I think we'll ever get.
&
I have learned a lot of scientific common sense and scientific thinking\\

\end{longtable}
\end{center}

\begin{center}
\footnotesize
\begin{longtable}{ l | c | c }
\caption{Average course scores by demographic groups. While our course design did not explicitly target students’ grades or performance, we observe no differences in average scores across demographic groups for gender (male and female) and culture/ethnicity (non-underrepresented minorities [non-URM] and underrepresented minorities [URM]).\label{tab:edpaper_grades}} \\

\hlineB{2} \multicolumn{1}{c|}{\textbf{Student Identity}} & \multicolumn{1}{c|}{\textbf{Average Cumulative}} & \multicolumn{1}{c}{\textbf{Welch's $t$-Test}} \\
\multicolumn{1}{c|}{\textbf{(Self-Reported)}} & \multicolumn{1}{c|}{\textbf{Course Score}} & \multicolumn{1}{c}{\textbf{$p$ value}}\\ \hlineB{2} 
\endfirsthead



\hlineB{2}
\endlastfoot
       
            Male (N = 10) & 84.3\% & \multirow{2}{*}{0.877} \\
            Female (N = 10) & 85.7\% & \\ \hline
 
            Non-URM (N = 16) & 84.8\% & \multirow{2}{*}{0.915} \\
            URM (N = 4) & 85.8\% & \\

\end{longtable}
\end{center}

\clearpage
\begin{center}
\footnotesize
\setstretch{1.15}
\setlength\tabcolsep{4pt}
\begin{longtable}{m{0.4\columnwidth}| c | c | c | c}
\caption{Survey items from the TSSI \citep{Cobern_TSSI} used in our pre- and post-course survey; see Sec.~\ref{sec:edpaper_TSSI} for a more detailed description. The second column indicates whether an item has ``reverse polarity'', i.e., if a student agrees with our course goals, they would respond with “strongly disagree”. Here we report data from 13 students who took both of the surveys. The survey's Likert scale is coded such that ``strongly disagree'' = -2, ``disagree'' = -1, ``neutral'' =  0, ``agree'' = 1, and ``strongly agree'' = 2. The final column is the difference in the average between the post- and pre-course survey results. The last row 
compares students' average pre- and post-course full survey responses; a Wilcoxon Signed-Rank Test was used to calculate the corresponding significance. \newline [Note: This table is split over 2 pages.]} \label{tab:edpaper_TSSI}\\

\hlineB{2} \multirow{2}{*}{\textbf{TSSI Prompt}} & \textbf{Reverse} & \textbf{Pre-Course} & \textbf{Post-Course} & \multirow{2}{*}{\textbf{$\Delta$}} \\ 
& \textbf{Polarity} & \textbf{Average} & \textbf{Average} & \\\hlineB{2}
\endfirsthead

\multicolumn{5}{c}%
{{\bfseries \tablename\ \thetable{} -- continued from previous page}} \\
\hlineB{2} \multirow{2}{*}{\textbf{TSSI Prompt}} & \textbf{Reverse} & \textbf{Pre-Course} & \textbf{Post-Course} & \multirow{2}{*}{\textbf{$\Delta$}} \\ 
& \textbf{Polarity} & \textbf{Average} & \textbf{Average} & \\\hlineB{2}
\endhead

\hlineB{2} \multicolumn{5}{r}{{\textit{Continued on next page}}} \\\hlineB{2}
\endfoot

\hlineB{2}
\endlastfoot

No form of knowledge can be completely certain - not even scientific knowledge.	&&	0.15&	0.15&	0.00\\ \hline
 Science should be taught at all school grade levels.	& &	1.46&	1.38&	-0.08\\ \hline
All students should study science during the secondary school grade levels. & &		1.23&	0.92&	-0.31 \\\hline
Developing new scientific knowledge is very important for keeping our country economically competitive in today's world.	&&	1.31&	1.31&	0.00\\\hline
A person can be both religious and scientific.	&&	1.23&	1.23&	0.00\\\hline
It is equally important for a person to have scientific knowledge and an appreciation for the arts.	&&	1.38&	1.46&	0.08\\\hline
Scientific knowledge is useful for only a few people.&	R	&1.38&	0.54&	-0.85\\\hline
Scientific knowledge is useful in keeping our national economy competitive in today's world.	&&	1.31&	1.38&	0.08\\\hline
Scientific research is generally very important.&&		1.38&	1.46&	0.08\\\hline
Women are welcome in science just as much as men are.&&		1.08&	0.85&	-0.23\\\hline
African Americans and other minority people are just as welcome in the scientific community as are white men.	&&	1.00&	1.00&	0.00\\\hline
Science can contribute to our appreciation and experience of beauty.&&		1.46&	1.31&	-0.15\\\hline
Even at the university level all students should study at least some science.	&&	0.62&	0.92&	0.31\\\hline
Science is our best source of useful knowledge.	&&	0.54&	1.00&	0.46\\\hline
Human emotion plays no part in the creation of scientific knowledge.&&		-0.08&-0.08	&0.00\\\hline
Scientific explanations tend to spoil the beauty of nature.&	R	&1.31&	1.15&	-0.15\\
There are many good things we can do today because of scientific knowledge.	&&	1.54&	1.31&	-0.23\\\hline
Most people really do not need to know very much science.&	R&	0.92&	0.69&	-0.23\\\hline
The scientific community is mostly dominated by white men and is often unfriendly to minority people.	&R	&0.23&	-0.23&	-0.46\\\hline
Scientific knowledge is useful.	&&	1.69&	1.15&	-0.54\\\hline
The methods of science are objective.&&		0.31&	0.69&	0.38\\\hline
Science can help us preserve our natural environment and natural resources.	&&	1.62&	1.46&	-0.15\\\hline
Only a very few people really understand science.&	R&	0.69&	0.31&	-0.38\\\hline
Scientific knowledge tends to erode spiritual values.&	R&	0.46&	0.23&	-0.23\\\hline
Understanding science is a good thing for everyone.	&&	1.54&	1.38&	-0.15\\\hlineB{2.5}
\multirow{2}{*}{\textbf{Overall average}}	&&	\multirow{2}{*}{\textbf{0.63}}	& \multirow{2}{*}{\textbf{0.70}}	&\textbf{0.07}\\
&&&&\textbf{($p = 0.170$)}\\

\end{longtable}
\end{center}

\newpage
\setstretch{1.0}
\bibliographystyle{plainnat}
\bibliography{edpaper}

\end{document}